\documentclass{osa-article}

\journal{oe}

\usepackage{color}

\DeclareUnicodeCharacter{2009}{\,} 

\begin{document}

\title{High speed imaging of spectral-temporal correlations in Hong-Ou-Mandel interference}

\author{Yingwen Zhang,\authormark{1} Duncan England,\authormark{1,*} Andrei Nomerotski,\authormark{2}  and Benjamin Sussman\authormark{1,3}}

\address{\authormark{1}National Research Council of Canada, 100 Sussex Drive, Ottawa ON Canada, K1A0R6\\
\authormark{2}Physics Department, Brookhaven National Laboratory, Upton, NY 11973, USA\\
\authormark{3}Department of Physics, University of Ottawa, Ottawa, Ontario, K1N 6N5, Canada}
\email{\authormark{*}duncan.england@nrc-cnrc.gc.ca} 


\begin{abstract}
In this work we demonstrate spectral-temporal correlation measurements of the Hong-Ou-Mandel (HOM) interference effect with the use of a spectrometer based on a photon-counting camera. This setup allows us to take, within seconds, spectral temporal correlation measurements on entangled photon sources with sub-nanometer spectral resolution and nanosecond timing resolution. Through post processing, we can observe the HOM behaviour for any number of spectral filters of any shape and width at any wavelength over the observable spectral range. Our setup also offers great versatility in that it is capable of operating at a wide spectral range from the visible to the near infrared and does not require a pulsed pump laser for timing purposes. This work offers the ability to gain large amounts of spectral and temporal information from a HOM interferometer quickly and efficiently and will be a very useful tool for many quantum technology applications and fundamental quantum optics research.
\end{abstract}

\section{Introduction}
The interference of two photons on a beamsplitter (BS), the Hong-Ou-Mandel (HOM) effect~\cite{Hong1987}, is one of the most defining features of quantum mechanics and lies at the heart of many technology applications requiring multi-photon interference including quantum communications, quantum computing, quantum state engineering and precision measurements ~\cite{Sangouard2011,Acin2007,Knill2001,Zhang2016,Zhao2014,Vitelli2013,Lyons2018,Lyons2018b,Bouchard2021}. The characteristic signature of HOM interference is a `dip' in two-fold coincidences in the output ports of the beamsplitter. It has become synonymous with HOM interference, but in reality this dip is only a single-mode representation of a far more complex multi-mode effect. Recent demonstrations of the HOM effect in spatial-temporal~\cite{Devaux2020}, and spectral-temporal~\cite{Gerrits2015,Jin2015,Thiel2020,Chen2020} modes have revealed the complex multi-dimensional nature of HOM interference. Spectrally resolving the HOM interference of light-atom interactions have also been recently demonstrated~\cite{Lipka2021}.

Here we use a two-photon spectrometer based on a state-of-the-art photon-counting camera~\cite{Ianzano2020,Nomerotski2019,Nomerotski20202} to spectrally resolve the HOM effect with high spectro-temporal resolution. Each output port of the beasmplitter (BS) is spectrally binned into 256 pixels and coincident detection is performed in all $256\times256=65536$ joint spectral modes simultaneously as the arrival time between the photons is scanned. In this way, a full spectro-temporal characterization of the HOM effect can be performed within just a few minutes. Conventional single photon cameras, such as electron-multiplying and intensified charge-coupled device (EMCCD and ICCD) cameras, limited by their frame rate, will require hours if not days to perform the same measurement~\cite{Jachura2015,Sun2019} and are also prone to stability issues within the long data taking period. It is also well known that the shape and width of the HOM interference dip is closely related to the photon spectrum and type of spectral filter used to perform the experiment~\cite{Ou2007,Liu2012,Lingaraju2019}. Conventionally, investigating the spectral response of HOM interference may require the use of many different spectral filters or the use of pulse shapers. However, with our system, all such spectral information can be obtained in just one data taking run. Through post-processing, we have the freedom to artificially add any number of spectral filters with any wavelength, bandwidth and shape so long it is within the spectral range of the spectrometer. Compared to using a fiber-based spectrometer~\cite{Gerrits2015,Jin2015}, or a chirped fiber Bragg grating~\cite{Davis2017,Thiel2020}, our camera based spectrometer can operate in a much wider spectrum from around 200~nm to 900~nm depending on the spectral response of the image intensifier used. We are also not limited to analysing photon pairs generated from a pulsed laser pump source, often required for timing purposes, thus giving our system more versatility.    

\section{Theory}

The spectral correlation between the photon pairs generated from the process of spontaneous parametric down-conversion (SPDC) can be displayed in a joint spectral intensity (JSI) plot. Theoretically this can be determined by multiplying the phase-matching function of the crystal and the pump envelope function. The phase matching function for a ppKTP crystal of length $L$ with Type-0 phase matching is given by 
\begin{equation}
    h(\omega_1,\omega_2) = \text{sinc}[L\Delta k(\omega_1,\omega_2)]
    \label{phase}
\end{equation}
with
\begin{equation}
    \Delta k(\omega_1,\omega_2)= \frac{1}{2\pi c} \left[n_e(\omega_p)\omega_p - n_e(\omega_1)\omega_1  - n_e(\omega_2)\omega_2\right]  - \frac{1}{\Lambda_0},
    \label{k}
\end{equation}
where $\omega_p$, $\omega_1$ and $\omega_2$ are the frequency of the pump and photon pairs respectively. $n_e(\omega)$ is the extraordinary refractive index of the crystal, $\Lambda_0$ is the grating period of the crystal and $c$ the speed of light in vacuum.

Assuming a Gaussian shaped pump, the pump envelope function is given by
\begin{equation}
    \alpha(\omega_1,\omega_2) = \exp\left[-\frac{\left(\omega_p-\omega_1-\omega_2\right)^2}{2\sigma_p^2}\right],
    \label{pump}
\end{equation}
with $\sigma_p$ the pump bandwidth. 

Finally the JSI is given by 
\begin{equation}
    \Phi(\omega_1,\omega_2)=|\alpha(\omega_1,\omega_2)h(\omega_1,\omega_2)|^2.
    \label{JSI}
\end{equation}

Now with an expression for the JSI we can use it to calculate the shape of the HOM dip. As shown in \cite{Hong1987,Ou2007}, the intensity correlation function for the HOM interference with multi-spectral modes is given by
\begin{equation}
    G^{(2)}(t_1,t_2,\Delta z) \propto \left|\int d\omega_1d\omega_2\Phi(\omega_1,\omega_2) \left(T e^{i\omega_1\Delta z/c} - Re^{i\omega_2\Delta z/c} \right)e^{-it_1\omega_1-it_2\omega_2} \right|^2,
\end{equation}
with $\Delta z$ the path length difference, $T$ and $R$ the transmissivity and reflectivity of the BS. 

Since the detector timing resolution is much longer than the coherence time of the SPDC photons, which is approximately the inverse of the spectral filter bandwidth, the two photon coincidence count is then proportional to the time average of $G^{(2)}(t_1,t_2)$ 
\begin{align}
    N_c(\Delta z) &\propto \int_\infty^\infty dt_1dt_2 G^{(2)}(t_1,t_2,\Delta z) \nonumber\\
        &\propto \left|\int d\omega_1d\omega_2 \Phi(\omega_1,\omega_2) \left(T e^{i\omega_1\Delta z/c} - Re^{i\omega_2\Delta z/c} \right)\right|^2.
    \label{dip}
\end{align}

A useful observation here is the absolute square of the integrand in Eq.~\ref{dip}, 
\begin{equation}
    \left|\Phi(\omega_1,\omega_2) \left(T e^{i\omega_1\Delta z/c} - Re^{i\omega_2\Delta z/c} \right)\right|^2,
    \label{HOMJSI}
\end{equation}
will be the JSI measured between the two output ports of the BS in HOM interference. Oscillations in the intensity will be observed with a frequency of $2\Delta z/c$.

When spectral filters with spectral shape of $f(\omega)$ are placed in the system, a small modification is required in the JSI, $\Phi(\omega_1,\omega_2)\rightarrow\Phi'(\omega_1,\omega_2)$, where
\begin{equation}
    \Phi'(\omega_1,\omega_2) = f(\omega_1)f(\omega_2)\Phi(\omega_1,\omega_2).
    \label{filter}
\end{equation}

\section{Experimental Setup}

\begin{figure*}[htbp]
	\centering \includegraphics[width=1\textwidth]{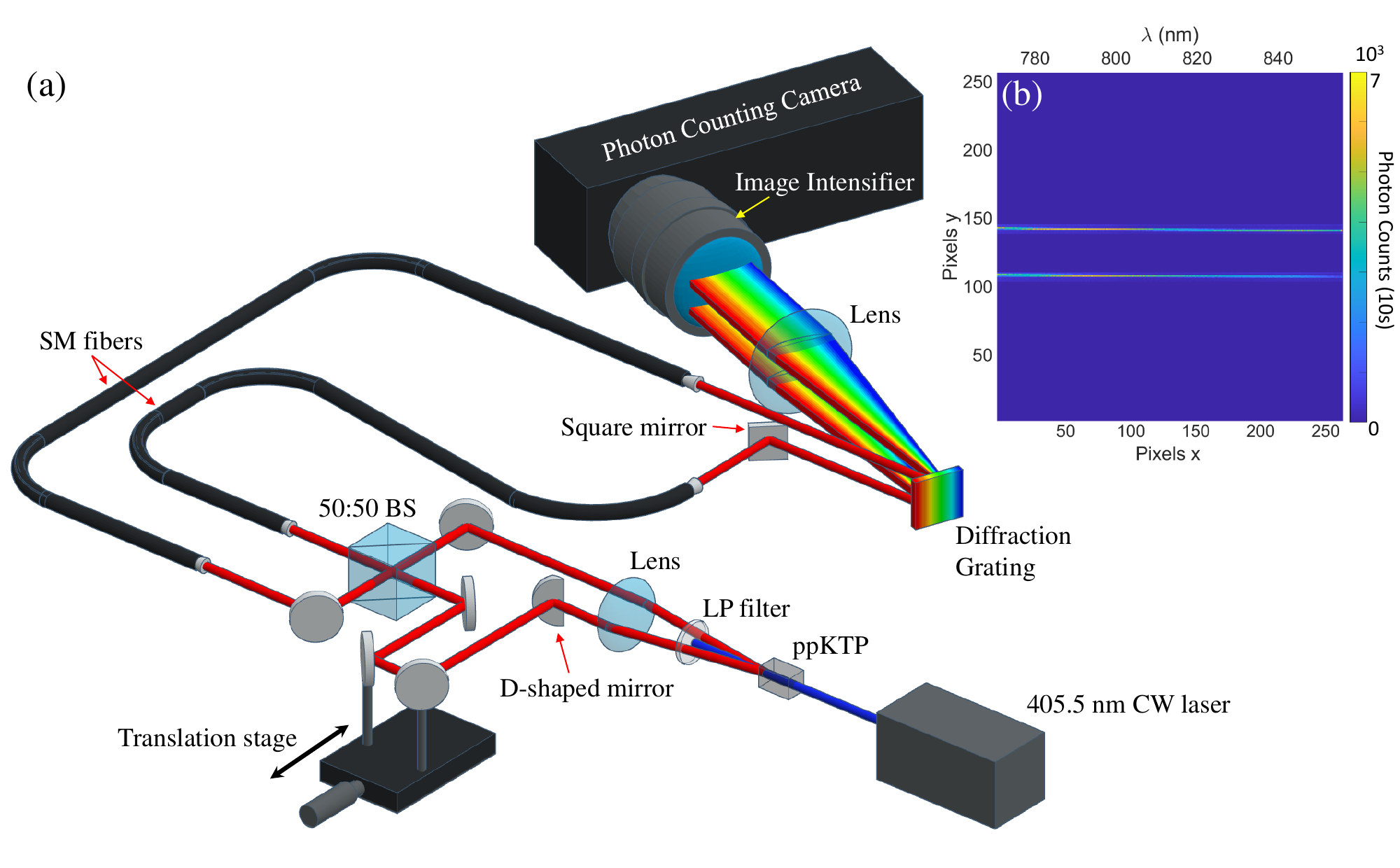}
	\caption{(a) Time and wavelength correlated photon pairs are generated through SPDC by a 405.5\,nm CW laser pumping a nonlinear ppKTP crystal. A long-pass~(LP) spectral filter is used to filter out the pump laser. The generated signal and idler photons are then sent on different paths and undergo HOM interference at a 50:50 BS before being coupled into single-mode~(SM) optical fibers. A motorized translational stage is used to control the arrival time difference between the two photons at the BS. The photons are vertically displaced upon exiting their respective fibers and they are diffracted off a diffraction grating which decomposes the photons into their spectral compositions. The diffracted photons are then focused onto the camera by a lens, placed at 1 focal length from the grating and camera. (b) Image of the twin beam spectrum captured on the camera with photons accumulated over 10 seconds.}
	\label{Fig.1}
\end{figure*}

The experimental setup is shown in Fig.~\ref{Fig.1}(a). A 200\,mW, 405.5\,nm continuous wave~(CW) laser (Toptica iBeam smart) with a bandwidth (FWHM) of 1.2\,nm is used to pump a 1\,mm long Type-0 periodically poled potassium titanyl phosphate~(ppKTP) crystal to generate photon pairs through the process of SPDC. A 10\,cm focal length lens is placed at 1 focal length after the crystal to collimate the SPDC beam and bring the far field of the the photon pairs onto a D-shaped mirror, which taking advantage of the photon pair's momentum anti-correlation, splits the photons into separate paths. The photon pairs are then sent into a 50:50 BS (actual T:R ratio is 57:43) to undergo HOM interference. The difference in arrival time between the photon pairs at the BS is adjusted through a motorized delay line constructed in one of the paths. Upon exiting the BS, the photons are coupled into single mode optical fibers which are then fed into the customized spectrometer~\cite{Johnsen2014,Zhang2020,Svihra2020}.

Within the spectrometer, the exit of the fibers are set at a slightly different height with respect to each other and the two beams are diffracted off a diffraction grating with the first diffraction order focused onto the photon-counting camera (TPX3CAM~\cite{ASI}) using a 2~inch diameter, 10\,cm focal length lens placed at 1 focal length from the grating and camera. The two beams are made to not propagate perfectly in parallel so to prevent the lens from focusing both spectra onto the same strip on the camera. The timing and spectral information of the photons are registered by the camera from which we are able to deduce the temporal and spectral correlations between the photons. Single photon sensitivity of the camera is provided by the attached image intensifier which has a quantum efficiency of approximately 20\% in the spectral range of 500-850\,nm~\cite{cricket}. A raw image of the twin beam spectrum captured on the camera is seen in Fig.~\ref{Fig.1}(b). 

As a single photon is converted into a flash of light by the image intensifier, a cluster of pixels will often be illuminated on the camera. Such a cluster has to be regrouped into a single event through a detection and centroiding algorithm~\cite{Ianzano2020}. After this correction, a spectral resolution of approximately 0.7\,nm is obtained over a spectral range of 770-854~nm. Due to the signal strength (photon flux) dependence of the camera's discrimination threshold crossover time~\cite{Ianzano2020,Zhao2017}, a timing correction must also be performed, after which a timing resolution of approximately 6\,ns is achieved. More details on the camera and it's calibration can be found in the Appendix.

\section{Results}

\begin{figure*}[htbp]
	\centering \includegraphics[width=1\textwidth]{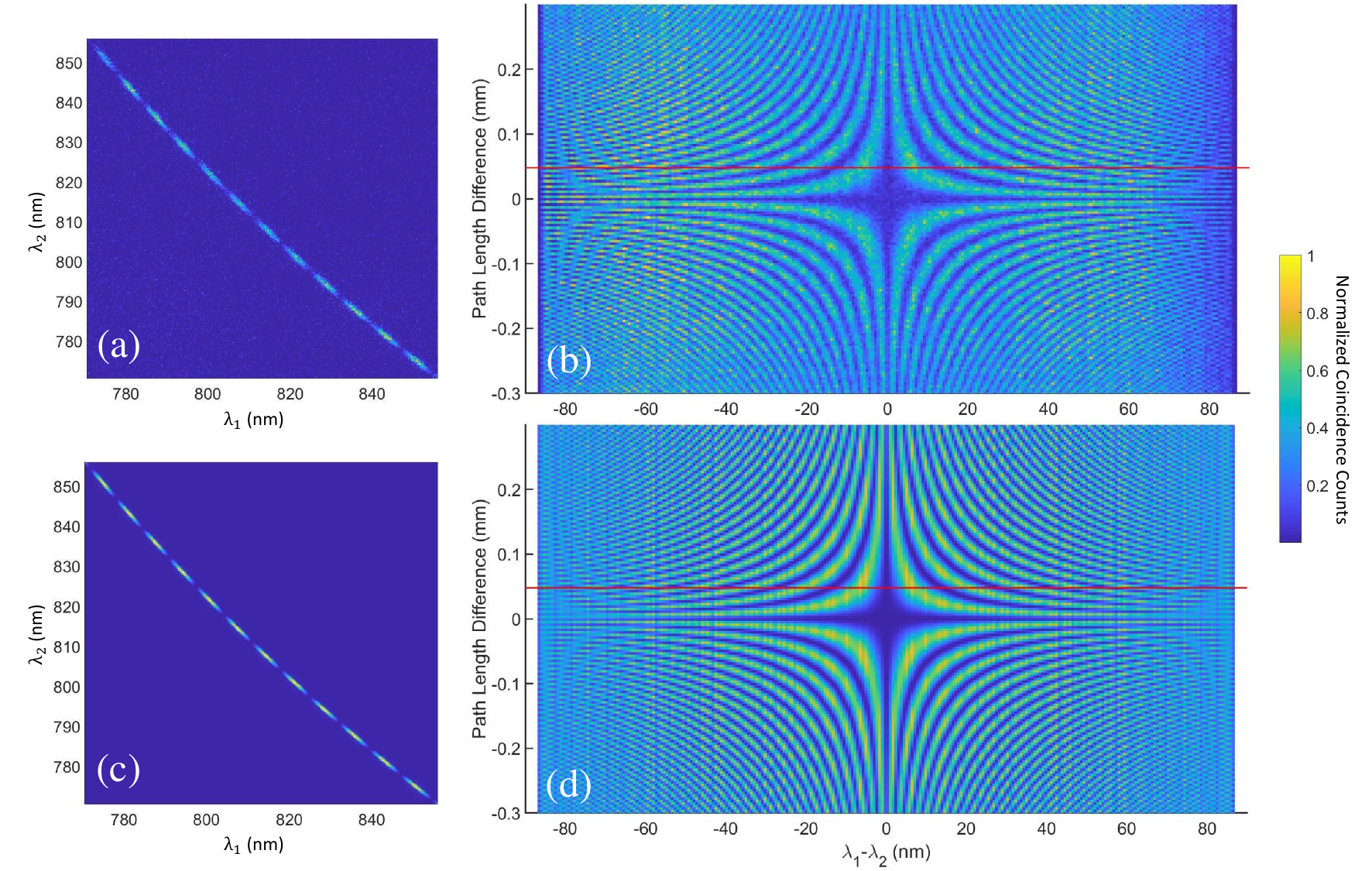}
	\caption{(a) A typical JSI measured after the BS at a path length difference of $\Delta z=0.048$\,mm. (b) Cross-section along the bright central band of the experimentally measured JSI as a function of the path length difference $\Delta z$ with $\lambda_1-\lambda_2=0$ the degenerate wavelength. The cross-section of (a) highlighted by the red line. The corresponding theoretical JSI are shown in (c) and (d). A total of 151 JSI slices were taken in (b) between $\Delta z=-0.3$\,mm and $0.3$\,mm, the data acquisition time is 10~seconds with a total of $\sim10^4$ coincidence events for each JSI slice. Note that the coincidence counts here are not the total coincidence $N_c(\Delta z)$ given by Eq.~\ref{dip} but the coincidence for each wavelength pair given by Eq.~\ref{HOMJSI}.}
	\label{Fig.2}
\end{figure*}

Experimentally, through looking for time correlated events between the two spectral bands seen in Fig.~\ref{Fig.1}(b) and then by recording the two corresponding x-pixel/wavelength of the correlated photons, a JSI plot can be made. A typical JSI experimentally measured after the BS at a path length difference of $\Delta z=0.048$\,mm is shown in Fig.~\ref{Fig.2}(a). When averaging the coincidence events across the entire spectrum at this $\Delta z$, as done in conventional time-only correlation measurements, no photon bunching can yet be observed to identify the presence of HOM interference. Therefore, to observe HOM interference at larger $\Delta z$ with only time correlation measurements, narrow spectral bandpass filter will need to be used. However, with the addition of spectral-correlation measurements, the presence of HOM inteference can be identified at a much larger $\Delta z$ through the observation of the interference fringes in the JSI plot. The cross-section along the bright central band of the experimentally measured JSI as a function of $\Delta z$ is shown in Fig.~\ref{Fig.2}(b). We see fast oscillations in the coincidence rate when away from $\Delta z=0$ and no coincidences at $\Delta z=0$. The corresponding theoretical JSI, calculated from the integrand of Eq.~\ref{HOMJSI}, are shown in Fig.~\ref{Fig.2}(c) and (d). Very good agreement is seen between the experimental results and theory.

In order to calculate the theoretical JSI, we first calculated the phase matching function (Eq.~\ref{phase} and \ref{k}) with $\lambda_p=2\pi c/\omega_p=405.5$\,nm, $\Lambda_0 = 3.425$\,$\mu$m and $L=1$\,mm. $n_e(\omega)$ is obtained using the Sellmeier equation given in \cite{Manjooran2012} for a temperature of $24^\circ \text{C}$. Using the 1.2\,nm FWHM bandwidth of the laser and assuming a Gaussian pump envelope function (Eq.~\ref{pump}) gives a $\sigma_p$ of 0.5\,nm. 

\begin{figure*}[htbp]
	\centering \includegraphics[width=1\textwidth]{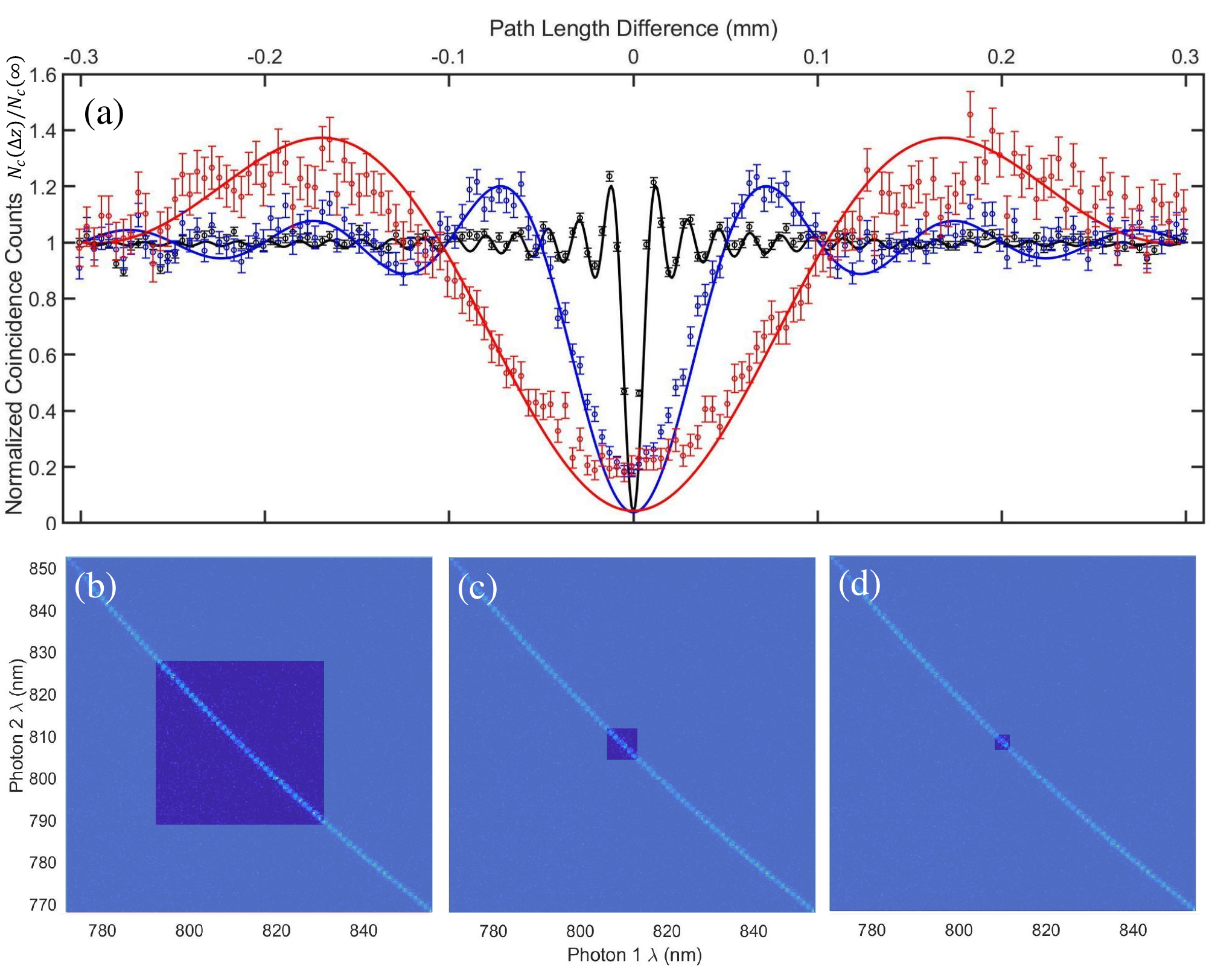}
	\caption{(a) The HOM interference dip for top-hat shaped spectral filters centered on the degenerate wavelength of 811\,nm with a bandwidth of 40\,nm (black), 7\,nm (blue) and 3\,nm (red). Circles are experimental data and the corresponding solid line with the same color is the predicted shape based on theoretical calculation. (b)-(d) highlights the corresponding region in the full JSI for the three filter bandwidths 40\,nm, 7\,nm and 3\,nm, respectively}. The displayed JSI is taken at path length difference of $\Delta z=0.3$\,mm.
	\label{Fig.3}
\end{figure*}

Since the measured JSI in Fig.~\ref{Fig.2} contains all the coincidence events across the full measurable spectrum of the spectrometer, one can through post-processing, place virtual spectral filters over the measured JSI to look at the changes in coincidence events in any spectral region of interest. In Fig.~\ref{Fig.3}, we show the HOM dip for top-hat shaped spectral filters of various bandwidth (3, 7 and 40\,nm) centered on the degenerate wavelength of 811\,nm extracted from the HOM dip scan data taken with no physical spectral filters used. The shape of the HOM dip between theoretical calculation (through Eqs.~\ref{dip} and \ref{filter} using a BS T:R ratio of 57:43) and experimental data is shown in Fig.~\ref{Fig.3}(a).  Figure~\ref{Fig.3}(c)-(d) highlights the corresponding region in the full JSI, taken at $\Delta z = 0.3$\,mm, for the various filter bandwidth. The experimental dip did not fully reach the theoretical minimum is due to imperfections in the optics alignment.

\begin{figure*}[htbp]
	\centering \includegraphics[width=1\textwidth]{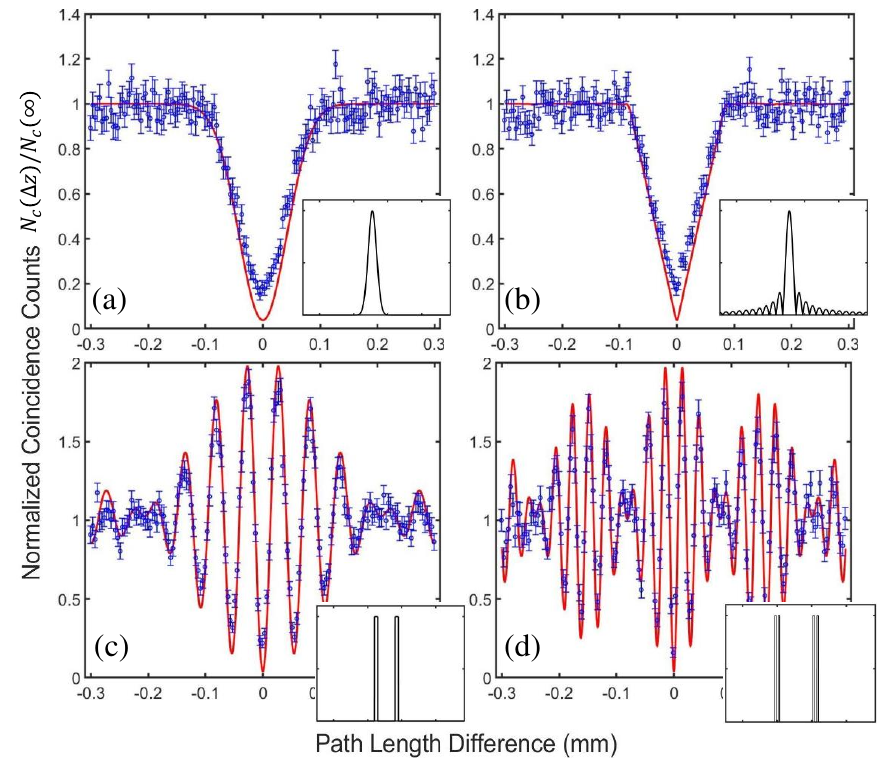}
	\caption{HOM interference dips for (a) a Gaussian shaped spectral filter centred on 811\,nm with a spectral bandwidth of 5\,nm (811/5\,nm),  (b) an absolute value Sinc function shaped 811/5\,nm spectral filter (see the Results section for the expression used for the Gaussian and Sinc function and their respective bandwidth)}, (c) dual band top-hat shaped filters centred on 805/2\,nm and 817/2\,nm and (d) quad band top-hat shaped filters centred on 799/1\,nm, 801/1\,nm 821.3/1\,nm and 823.4/1\,nm. Circles are experimental data and the solid line is the predicted shape based on theoretical calculation. The corresponding shape of the spectral filters are shown in the insets.
	\label{Fig.4}
\end{figure*}

Of course, one is also free to choose the shape and number of spectral filters to be placed and this is demonstrated in Fig.~\ref{Fig.4}. Figure.~\ref{Fig.4}(a) shows the HOM dip scan of a Gaussian shaped spectral filter where a Gaussian shaped HOM dip is observed. The expression used for the Gaussian filter is $f(\lambda) = \exp{\left[-2\left(\frac{\lambda-\lambda_c}{\Delta\lambda}\right)^2\right]}$, with the central wavelength of the filter $\lambda_c = 811$\,nm and the spectral bandwidth $\Delta\lambda = 5$\,nm. In Fig.~\ref{Fig.4}(b), an absolute valued Sinc function filter is used to give a triangular shaped dip. The expression used for this filter is $f(\lambda) = \left|\sqrt{\text{sinc}\left[\frac{4\left(\lambda-\lambda_c\right)}{\Delta\lambda}\right]}\right|$ with a filter central wavelength of $\lambda_c = 811$\,nm and a spectral bandwidth $\Delta\lambda = 5$\,nm. To apply these two filters on the experimental data, coincidence events were randomly discarded in post-processing with a probability based on the amplitude of the applied filter. In Fig.~\ref{Fig.4}(c), a dual band top-hat filter centered on 805/2nm (central wavelength 805\,nm with 2\,nm bandwidth) and 817.1/2\,nm is used  where oscillations are seen in the HOM dip. Finally, in Fig.~\ref{Fig.4}(d), a quad band top-hat shaped shaped filters centred on 799/1\,nm, 801/1\,nm 821.3/1\,nm and 823.4/1\,nm is shown, in which the combination of narrow filters and small central wavelength difference results in a beating pattern in the HOM dip oscillation. 

\begin{figure*}[htbp]
	\centering \includegraphics[width=1\textwidth]{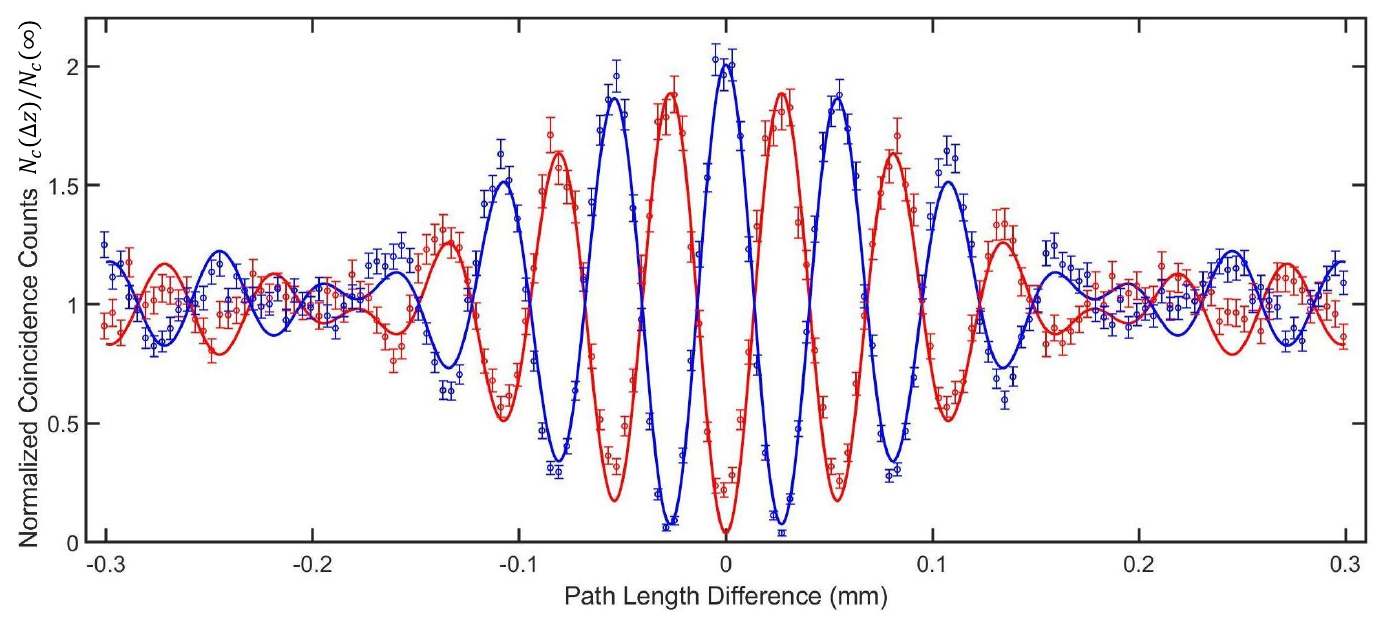}
	\caption{The HOM interference observed in the same output channel of the BS (blue) and observed between separate output channels of the BS (red) for dual band top-hat shaped filters centred on 805/2 nm and 817/2 nm. Circles are experimental data and the corresponding solid line with the same color is the predicted shape based on theoretical calculation. The blue data shown are the sum of the coalescence events from both output channels.}
	\label{Fig.5}
\end{figure*}

It is also possible to observe photon coalescence in HOM from the same data set by looking for coincidences in the spectrum of the same beam instead of between the two beams\cite{DiGiuseppe2003,Nomerotski20202}. However, due to the requirement to correct for clustering of events caused by the image intensifier, as discussed in the previous section, coincidence events that occur approximately in a 10 pixel radius around the degenerate wavelength of 811\,nm will all be regrouped into a single event. This results in the inability to obtain an accurate HOM `peak' for single band spectral filters centered around the degenerate wavelength of 811\,nm. However, for non-degenerate wavelengths, as seen in Fig.~\ref{Fig.5}, accurate HOM interference shape for photon coalescence can still be obtained.

\section{Discussion}

In this work we demonstrated measurement of the spectral and temporal correlations in HOM interference using a camera based spectrometer. The spectral response to HOM interference for the full operating spectrum of the spectrometer, between 770-854\,nm, is obtained with a spectral resolution of 0.7\,nm. From this, the HOM interference behaviour at any specific spectral region/s within the measured spectrum can be extracted. Thus, with this spectrometer, full spectral and temporal correlation data, with sub-nanometer spectral and nanosecond temporal resolution, can be obtained within seconds and a full HOM dip/peak scan can be performed in minutes. The system is also versatile in that it can be used with SPDC sources pumped by any type of laser and can work in any wavelength that an image intensifier is sensitive in. All these technical improvements combined, especially in speed, opens up a whole new avenue on what is practicable in quantum technology applications and experiments. Some examples include spectral multiplexing in quantum repeaters, obtaining spectral information of targets through a single shot measurement in quantum sensing and swiftly characterizing spectral-temporal properties of single photon sources. Presently, high quantum efficiency commercial intensifiers can be found in the wavelength range of 200-900\,nm, thus, for related experiments and technologies utilizing telecom wavelength photons (1550\,nm), a fiber spectrometer will still be required. However, if an intensifier or other similar photon amplification devices, sensitive in the telecom wavelength, were to be developed in the future, they can be readily integrated into this camera spectrometer.

Improvements to the spectral resolution can be achieved with a better designed spectrometer by either illuminating more lines on the diffraction grating or using a grating with more groves and using a different focusing lens to fill the camera with a narrower spectral region. Additionally, through predicting the initial position of the photon using a centroiding algorithm and subdividing each camera pixel, a higher spatial resolution than the physical number of pixels on the camera can also be achieved\cite{Kim2020}. The temporal resolution can also be improved to sub-nanoseconds with the use of a trigger signal from the multi-channel plates of the image intensifier \cite{Nomerotski2020} or, as proposed in the same work, using of a streaking tube in place of the image intensifier. The measurement speed is only limited by the photon pair production rate and the overall detection efficiency of the system, a brighter source and more efficient optical elements (mostly the diffraction grating) will further improve the overall measurement speed.

\section*{Appendix}
\subsection*{Camera Specifications and Working Principle}
The TPX3Cam is a state-of-the-art event based photon-counting camera system in that a time stamp can be applied to every photon detected on each pixel. The raw data output is thus a string of numbers listing the time and pixel number for every detected photon instead of individual frames as in conventional camera systems. The camera has a silicon based sensor with $256\times256$ individual pixels, each with a timing resolutions of 1.6\,ns. This timing capability on individual pixels comes at the cost of a large pixel size of $55\times55$\,$\mu$m$^2$ to accommodate the hundreds of transistors on each pixel. With the transistors and electronics attached under the pixels, a camera fill factor of near 100\% is achieved. Due to interference from the large number of electronics attached, the pixel noise is too high to be single photon sensitive, however with a light flash bright enough to cross the camera's discrimination threshold, an additional property, the intensity of the light flash can be roughly estimated as a time-over-threshold (TOT) - the total time taken for the signal to cross the discrimination threshold and decay back down again. Single photon sensitivity is provided by an attached image intensifier. In the image intensifier, an electron is ejected by a photon on the photocathode and this single electrons is then multiplied into a cascade of electrons through the microchannel plate (MCP). The electrons then hits the phosphor screen which produces a flash of light that is bright enough to be finally registered by the camera.

As already briefly mentioned in the main text, a cluster of pixels will often be illuminated by the flash of light. Thus, such a cluster has to be regrouped into a single event through a detection and centroiding algorithm. In contrary to expectations, this cluster detection and centroiding algorithm can actually be used to improve upon the camera spatial resolution to beyond the physical number of pixels. With an average cluster size of around 7 pixels in diameter (this number will vary between systems depending on the gain of the intensifier, with larger gains producing more photons thus a larger cluster) and using the TOT as a weighting factor, the photon location can be pinpointed to an area smaller than the physical size of the pixels as demonstrated in~\cite{Kim2020}. As this will require additional computing time, we did not go beyond the pixel level with our centroiding algorithm. As the photon flux will vary between each pixel in a cluster covered by a light flash, and the camera’s discrimination threshold crossover time (time-of-arrival TOA) depends on the signal strength detected on each pixel, a timing correction must also be performed. Using the fact that each cluster came from the same photon and should have the same TOA, one can extrapolate a relation between the signal strength (TOT) and the difference in TOA from analysing tens of thousands of clusters~\cite{Ianzano2020,Zhao2017}. From this information, a timing correction is applied and a timing resolution of approximately 6\,ns can be achieved. Though still unable to reach the camera resolution of 1.6\,ns, without this timing correction step, the timing resolution will be worse by an order of magnitude.  

The pixel dead time on the camera is approximately 1\,$\mu$s, however, given that the photon count rate for this experiment is around 100\,kHz and is spread across hundreds of pixels, chances of having multiple hits on the same pixel within this 1\,$\mu$s time window is extremely low, there is negligible effect on our data. In the image intensifier, the MCP will require time to recover every time it amplifies an electron.  The time of the pore paralysis for standard MCPs is in the order of 1\,ms, which depends on the gain setting and MCP resistance. When an electron is amplified, generally more than one MCP channel is affected and charges held in neighboring channels are also utilized, but given that there are in general $\sim 10^5$-$10^6$ channels on the MCP, again, the probably of having multiple hits on a small cluster of channels within the 1\,ms time window is also very low. After some testing, we estimate that to saturate our intensifier will require a photon detection rate of approximately 50\,MHz when spread evenly across the whole intensifier, for the area of one of our spectral bands at about $2\times 256$ pixels, this limit is around 400\,kHz. The count rate for the experiment is still 4 times below this limit, thus should not be much affected by the intensifier dead time.

\subsection*{Calibration of Spectral Resolution}
The spectrometer is characterized by comparing the locations of interference fringes measured in the experiment to that calculated through theory as seen in Fig.~\ref{Fig.2}.  Three parameters, the central wavelength of the signal and idler photon and the full spectral bandwidth were adjusted in the theoretical calculations until a good agreement between theory and experiment is reached, this gave the spectral range of our spectrometer. The spectral resolution is determined by looking for the maximum number of observable interference fringes on the camera.

\subsection*{Total Quantum Efficiency of Setup}
Through the ratio between the total photons detected vs. the total coincidences, the total photon detection efficiency of the setup is calculated to be approximately 1.25\%. This is a combined efficiency resulting from all components in the setup which we approximated to be 20\% coupling efficiency into the SM fibers, 50\% efficiency of the diffraction grating, 20\% efficiency of the camera, 75\% efficiency from a total of 7 silver mirrors (96\% efficiency each) used in each photon path and 90\% efficiency from the lenses, spectral filters and 50:50 BS.

\subsection*{Multi Photon Pair Events}
With the slow frame rate of standard CCD cameras, multi-pair events is a big issue and would often require gating and a low photon pair production rate to be able to reliably identify individual photon pair events.  Since the TPX3CAM is an event based camera system, no external timing sources, such as using a pulsed pump laser, are required. Photon pairs are identified using time–correlation measurements performed between multiple camera pixels instead of between a pair of single-photon avalanche diodes as in most SPDC experiments. Given the 6\,ns effective timing resolution of the camera and that we gated our coincidences with a 20\,ns time window, the chances of multi-pair events to occur within this time window for our count rate of 100\,kHz is rare. Adding to the fact that we also applied a spectral correlation condition on top of this~\cite{Zhang2020}, the probability for multi-pair events to occur is further reduced by around 2 orders of magnitude. However, in the unlikely event that a second photon (either from background or a double photon pair production at the crystal) is detected in temporal and spectral coincidence with an actual pair of SPDC photons, there would be no way to identify this and it has to be treated as noise.

\section*{Funding}
Defence Research and Development Canada

\section*{Acknowledgements}
The authors are grateful to Philip Bustard, Frédéric Bouchard, Khabat Heshami, Rune Lausten, Denis Guay, and Doug Moffatt for technical support and stimulating discussion. 

\section*{Disclosures}
The authors declare no conflicts of interest.

\section*{Data Availability Statement}
The raw data underlying the results presented in this paper are not publicly available at this time due to its size but may be obtained from the authors upon reasonable request.


\bibliography{HOMref}

\end{document}